# Managing Basis Risks in Weather Parametric Insurance: A Quantitative Study of Diversification and Key Influencing Factors

Hang Gao[1], Shuohua Yang[2], Xinli Liu[2]

**ABSTRACT:**

Weather parametric insurance relies on weather indices rather than actual loss assessments, enhancing claims efficiency, reducing moral hazard, and improving fairness. In the context of increasing climate change risks, despite growing interest and demand,, weather parametric insurance's market share remains limited due to inherent basis risk—the mismatch between actual loss and payout, leading to "loss without payout" or "payout without loss." This paper proposes a novel empirical research using Monte Carlo simulations to test whether basis risk can be managed through diversification and hedged like other risks. Key findings include: (1) Portfolio basis risk and volatility decrease as the number of contracts increases, (2) Spatial relationships significantly impact basis risk, with risk levels correlating with the ratio between insured location, weather station, and disaster footprint radius, and (3) Event severity does not significantly impact basis risk, suggesting that catastrophic disaster severity should not hinder parametric insurance development.

---

[1] Corresponding author, China Agricultural Reinsurance Co.,Ltd.
[2] School of Economics, Peking University

# Introduction

Weather parametric insurance is an innovative risk management tool that has seen rapid development over the past few decades. Unlike traditional insurance, where payouts are based on actual loss assessments, this type of insurance is designed with payout criteria based on specific weather indices. Payouts are triggered when an insurance index attaches a predefined threshold, providing coverage against property damage caused by weather-related disasters (Wenner et al., 2003). A key advantage of weather parametric insurance is its simplified claims process, which relies solely on pre-determined trigger conditions, eliminating the need for complex loss assessments and claims procedures. This significantly reduces insurance costs, mitigates moral hazard and adverse selection risks, and enhances the efficiency and fairness of insurance payouts (Barnett & Mahul, 2007).

For example, India's Pradhan Mantri Fasal Bima Yojana (PMFBY) program, launched in 2016, has seen an average annual increase of 37% in applicants from 2021 to 2023, demonstrating the critical role of weather parametric insurance (Press Information Bureau, 2023). In Kenya, the Kenya Livestock Insurance Program (KLIP), supported by the Kenyan government and the World Bank, provides livestock loss insurance to over 18,000 pastoralists, continuing to be a key component of Kenya's agricultural risk management strategy (World Bank, 2023). Additionally, the R4 Rural Resilience Initiative, co-founded by the World Food Programme (WFP) and Oxfam America, aims to enhance farmers' resilience to climate risks through the provision of weather parametric insurance and other risk management tools (World Food Programme & Oxfam America, 2023). The concept of weather parametric insurance is also increasingly applied in infrastructure protection, catastrophe bonds, and index-based parametric financial derivatives, such as Puerto Rico's parametric catastrophe bonds. The Caribbean Catastrophe Risk Insurance Facility (CCRIF) utilizes parametric insurance to assist 16 Caribbean and American governments in hedging hurricane risks (C. Adam & D. Bevan, 2020).

Despite the numerous advantages and growing interest in weather parametric insurance, its market share remains low (Stoeffler, 2020). One of the primary constraints to its widespread adoption is basis risk. Basis risk refers to the mismatch between the weather index-based payout and the actual loss experienced by policyholders, manifesting as either "loss without payout" or "payout without loss" (Turvey, 2001). The existence of basis risk has led some potential policyholders to question the effectiveness of weather parametric insurance, thereby affecting its market acceptance (Jensen, Barrett, & Mude, 2016).

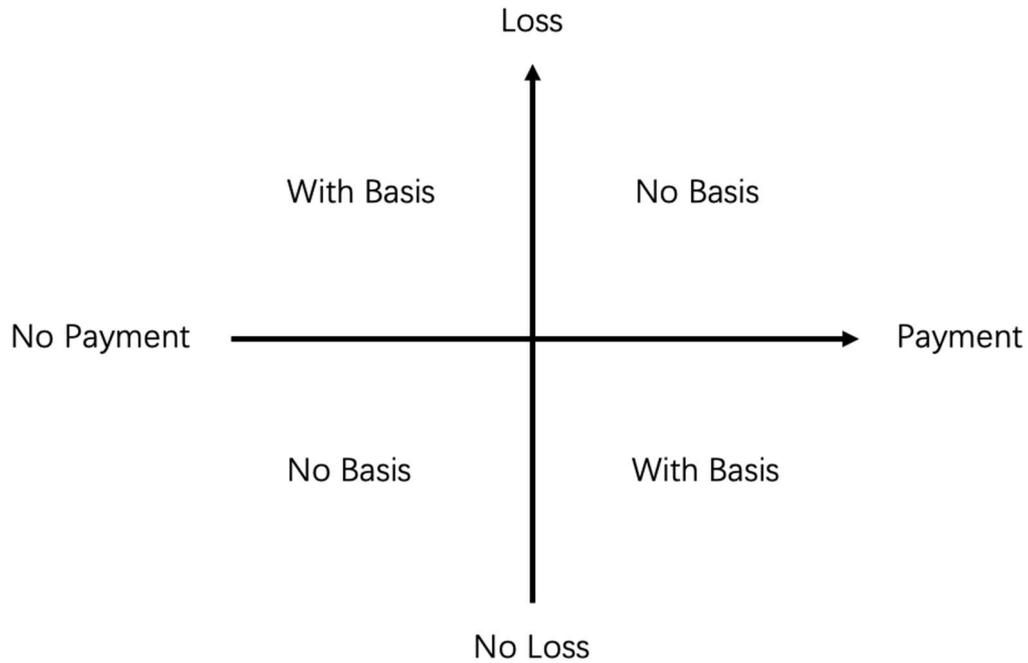

Figure 1 Quadrant Diagram of Basis Risk

Figure 1 illustrates a quadrant diagram of basis risk. By substituting index-based triggers for loss assessments, parametric insurance significantly simplifies the risk transfer mechanism. However, this simplification can result in discrepancies between insurance payouts and the actual losses experienced by policyholders, leading to potential basis risk. The relationship between basis risk, payouts, and losses is depicted in Figure 1.

Basis risk can be categorized into three main types: design basis risk, temporal basis risk, and spatial basis risk (Dalhaus et al., 2018). Numerous studies have explored these various types and investigated possible hedging approaches:

Design Basis Risk: This primarily stems from suboptimal trigger design and overly complicated payout models (Jensen, Barrett, & Mude, 2016). If the selected weather parameters are weakly correlated with actual loss, or if the insurance product's triggering mechanism is poorly designed, excessive or insufficient payouts may occur, increasing the insurer's risk (Miranda & Farrin, 2012). Researchers have proposed statistical solutions to improve parametric insurance performance. For instance,

Cesarini et al. (2021) demonstrated the potential of neural networks in identifying extreme weather events in the Dominican Republic, highlighting their ability to enhance classification accuracy and reduce basis risk. Nguyen-Huy et al. (2019) developed a Copula-based statistical model using the C-vine method to simulate joint insurance losses from drought events occurring simultaneously at different locations or continuously over different growing seasons, providing insights for pricing weather parametric insurance products.

Temporal Basis Risk: This risk arises from mismatches between the insurance tenure period and the actual crop production cycle, which are mainly influenced by climate change-induced alterations and plant growth characteristics. This risk can be mitigated through phenological observation networks or dynamic adjustments of insurance tenure periods using remote sensing vegetation indices (Afshar et al., 2021).

Spatial Basis Risk: This risk stems from the spatial misalignment between the location of the reference weather station, where observed weather measurements are used to determine if parametric insurance is triggered, and the location of insured exposure, where property losses actually occur. If the distance between the reference weather station and the insured exposure is too large compared to the hazard event footprint, basis risk becomes more pronounced. This is a distinctive feature of weather parametric insurance, which relies on external meteorological data, particularly ground-based weather station data, for both real-time observation for measuring triggers and historical data series for product design. This reliance on external data can be a potential hurdle in developing countries due to issues with data accuracy and continuity (Miranda & Farrin, 2012). Various studies have explored the use of more granular data, such as remote sensing indices, spatial dispersion, or interpolation techniques, as a possible solution to spatial basis risk (Ritter et al., 2014; Cao et al., 2015). Möllmann et al. (2019) assessed the application of three remote sensing vegetation health indices—Vegetation Condition Index (VCI), Temperature Condition Index (TCI), and Vegetation Health Index (VHI)—in weather derivatives in Germany, finding that VHI-based products significantly reduced basis risk, particularly in areas with sparse weather station networks.

Basis risk, an inherent consequence of the misalignment between actual insurance risk and risk proxies, cannot be completely eliminated through technical means. Existing research has primarily focused on design aspects, emphasizing improvements in index selection, model construction, and payout design. Efforts to reduce basis risk have included the use of finer or broader data, index filtering, optimization of triggering mechanisms, and the adoption of advanced modeling techniques. However, according to the literature, many of the proposed methods are too complex to be applied in practice, limiting their practical implications. Additionally, there has been a lack of clear quantitative definitions of basis risk to evaluate the effectiveness of management approaches. Furthermore, research addressing the effectiveness of diversification from an empirical business paradigm perspective—such as forming risk portfolios with multiple independent parametric insurance contracts—in mitigating basis risk is scarce. Overall, there is a lack of comprehensive quantitative analysis of the relationship between basis risk and its major influencing factors, making it challenging to fully elucidate the mechanisms and impact pathways of basis risk.

## Method

Assume the risk portfolio consists of several independent parametric insurance contracts. For any individual insurance contract: (a) The hazard footprint area is a circle with a radius $r$, where $r$ is a random number defined on $[0, R_{max}]$, and the centroid $[x_1, y_1]$ of the circle footprint is a random variable defined on $[0, 1]$; hazard severity within the footprint is uniform and represented by $s$ which is a random number defined on $[0, S_{max}]$ where $S_{max}$ is the maximum severity; (b) The location of the insured exposure is a fixed point with coordinates $[x_2, y_2]$ defined on $[0, 1]$; (c) The location of the reference weather station is a fixed point with coordinates $[x_3, y_3]$ defined on $[0, 1]$; (d) The parametric insurance product's trigger threshold $t$ is a fixed value for any individual contract and is less than $S_{max}$; (e) Any individual contract is assigned a uniform premium value of 1. The total premium of the risk portfolio is $m$, where n is the total number of contracts contained in the risk portfolio.

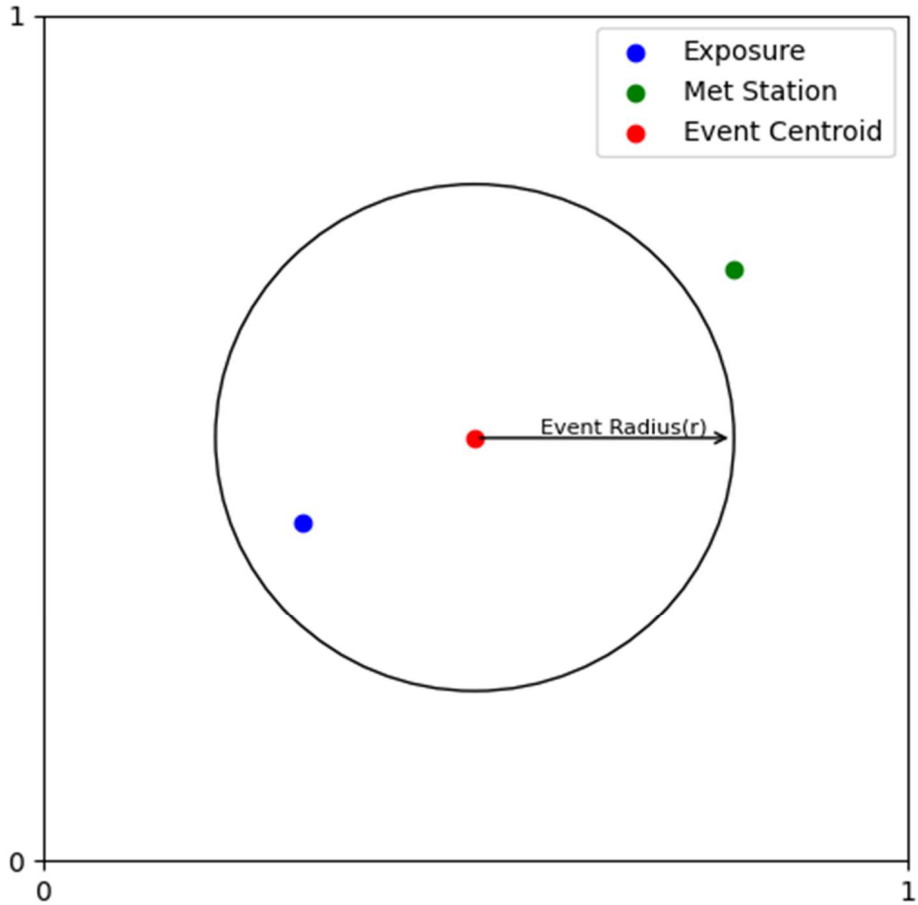

Figure 2 Schematic Diagram of Research

As illustrated in Figure 2, trigger conditions are categorized as follows:

(a) If event severity is greater than or equal to the threshold ($s \geq t$) and the weather station is within the event footprint area, the payout is triggered and equals 1; otherwise, the payout is 0. If event severity is greater than or equal to the threshold ($s \geq t$) and the insured exposure is within the event footprint area, a loss occurs and equals -1; otherwise, the loss is 0.

(b) If event severity is less than the threshold ($s < t$), neither payout nor loss is triggered, regardless of their spatial relation to the event footprint.

The spatial relationship among the event footprint, insured exposure, and reference weather station can be categorized into four scenarios:

    1. The event footprint simultaneously covers both the weather station and the insured exposure.

    2. The event footprint covers the weather station only, but not the insured exposure.

    3. The event footprint covers the insured exposure only, but not the weather

station.

4. The event footprint does not cover either the weather station or the insured exposure.

Based on these scenarios, basis risk is defined as follows:

(a) When a payout is triggered (payout = 1) and a loss is not triggered (loss = 0), basis risk is 1.

(b) When a payout is not triggered (payout = 0) and a loss is triggered (loss = -1), basis risk is -1.

(c) When both payout and loss occur simultaneously, they offset each other, resulting in a basis risk of 0.

(d) When neither payout nor loss occurs, basis risk is 0.

Table 1 Basis Risk Testing Model

Trigger Scenario

| | | **Trigger Condition Met** | **Trigger Condition Not Met** |
|---|---|---|---|
| event footprint | **Covers both weather station and insured object** | payout=1; loss=-1<br>no basis risk | payout=0; loss=0<br>no basis risk |
| | **Covers only the weather station, not the insured object** | payout=1; loss=0<br>basis risk present | payout=0; loss=0<br>no basis risk |
| | **Covers only the insured object, not the weather station** | payout=0; loss=-1<br>basis risk present | payout=0; loss=0<br>no basis risk |
| | **Neither the insured object nor** | payout=0; loss=0<br>no basis risk | payout=0; loss=0<br>no basis risk |

| | the weather station is covered |
|---|---|

In our Monte Carlo simulation, assuming a risk portfolio contains $m$ independent weather parametric insurance contracts, for the $i$-th ($i = 1,2,\ldots,m$) parametric insurance contract, associated variables are defined as follows:

Contract premium $P(i) = 1$;

Coordinates of the insured exposure: $(x_1(i), y_1(i))$;

Coordinates of the reference weather station: $(x_2(i), y_2(i))$;

Trigger threshold of the $i$-th contract: $t(i)$;

Distance between the weather station and the insured exposure:

$$d(i) = \sqrt{(x_1(i) - x_2(i))^2 + (y_1(i) - y_2(i))^2}.$$

For each parametric insurance contract, $n$ random simulation years are conducted. For each simulation year $j$ ($j = 1,2,\ldots,n$), associated random variables are defined as follows:

Coordinates of the hazard event centroid: $(x_3(i), y_3(i))$;

Hazard event radius: $r(i, j)$;

Hazard severity: $s(i, j)$.

Basis risk for the $i$-th contract in the $j$-th simulation year is denoted as $BR(i, j)$:

(a) If both the weather station and the insured object are within the disaster radius and $s(i,j) \geq t(i)$, both payout and loss occur, resulting in $BR(i, j) = 0$. If $s(i, j) < t(i)$, neither payout nor loss occurs, and $BR(i, j) = 0$.

(b) If neither the weather station nor the insured object is within the disaster radius, neither payout nor loss occurs, resulting in $BR(i, j) = 0$.

(c) If the weather station is within the event footprint but the insured exposure is not, and $s(i,j) \geq t(i)$, the payout equals 1 and the loss equals 0, resulting in $BR(i, j) = 1$. If $s(i, j) < t(i)$, both payout and loss are 0, and $BR(i, j) = 0$.

(d) If the insured exposure is within the event footprint but the weather station is not, and $s(i,j) \geq t(i)$, the payout equals 0 and the loss equals -1, resulting in $BR(i, j) = -1$. If $s(i, j) < t(i)$, both payout and loss are 0, and $BR(i, j) = 0$.

The basis risk level for the $i$-th contract is represented by the Annual Average Basis Risk per Premium Ratio (AABRP):

$$AABRP = \frac{Annual\ Average\ Basis\ Risk}{Premium}$$

$$AABRP(i) = \frac{\sum BR(i,j)}{nP(i)}, \quad j = 1,2,\ldots,n$$

The overall basis risk uncertainty for the *i-th* contract is represented by the standard deviation.

$$\sigma(i) = \frac{STDEV(BR(i,j))}{P(i)}$$

For a risk portfolio containing *m* independent contracts, the overall basis risk for the *j-th* simulation year is the sum of the basis risks of all *m* contracts:

$$BR'(j) = \sum(BR(i,j)), \quad i = 1,2,..m$$

The overall AABRP ratio for the risk portfolio is represented as $AABRP'$ which is the ratio of the average annual basis risk to the total premium of the risk portfolio:

$$AABRP' = \frac{average\left(\sum(BR'(j))\right)}{\sum P(i)}, \quad j = 1,2,..n, i = 1,2,...m$$

The overall basis risk uncertainty for the risk portfolio is represented by the standard deviation:

$$\sigma' = \frac{STDEV(BR'(j))}{\sum P(i)}, \quad j = 1,2,..n, i = 1,2,...m$$

## Analysis and Results

（1）**Basis risk of an individual contract and basis risk of a portfolio containing multiple contracts**

We assume a portfolio containing *m = 100* contracts and conduct a test with *n = 1000* simulation years. For each contract, the parametric trigger threshold is a fixed value randomly selected from a range within (0, 10]. For each simulation year, event severity is randomly selected from a range within [0, 20]. The *AABRP(i)* for individual contracts *( i = 1, 2, ..., 100 )* and the overall portfolio basis risk $AABRP'$ are calculated.

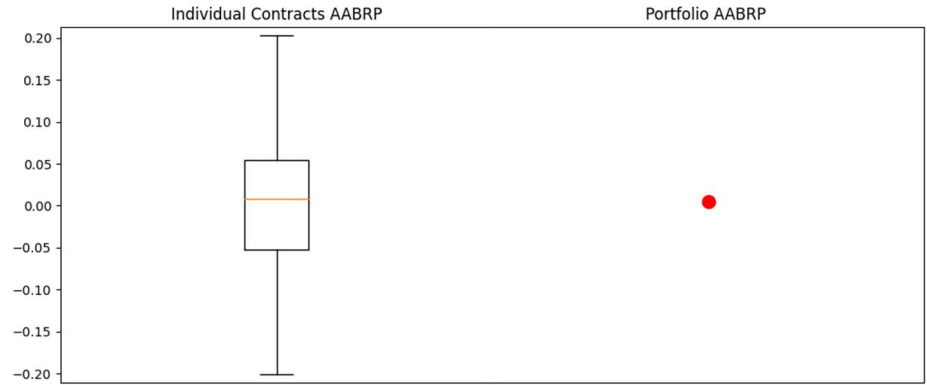

Figure 3 Comparison of AABRP Distribution for Individual Contracts and AABRP'
for Risk Portfolio

In the test, the AABRP for individual contracts ranges from -0.201 to 0.203, with the lower quartile at -0.053, the median at 0.007, and the upper quartile at 0.054. The overall portfolio AABRP' converges to 0, specifically 0.005, reflecting the hedging effect of basis risks given diversification.

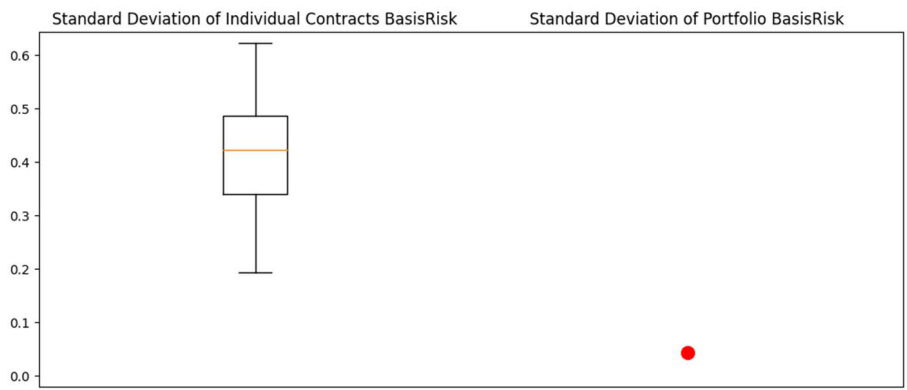

Figure 4 Comparison of Basis Risk Uncertainty between Individual Contracts and
Risk Portfolio

The basis risk standard deviation $\sigma(i)$ for any individual contract ranges from 0.118 to 0.589. The overall portfolio standard deviation $\sigma'$ converges to 0, specifically 0.041, indicating a significant reduction in the volatility of basis risks. This demonstrates that diversification could potentially improve the manageability of basis risks in a portfolio.

In another set of tests, each based on 1000 simulation years, while keeping other simulation parameters constant, we increase the number of contracts in the risk portfolio from 1 to 500. The absolute value of the overall portfolio basis risk AABRP' decreases and evidently converges toward 0, as shown in Figure 5(a).

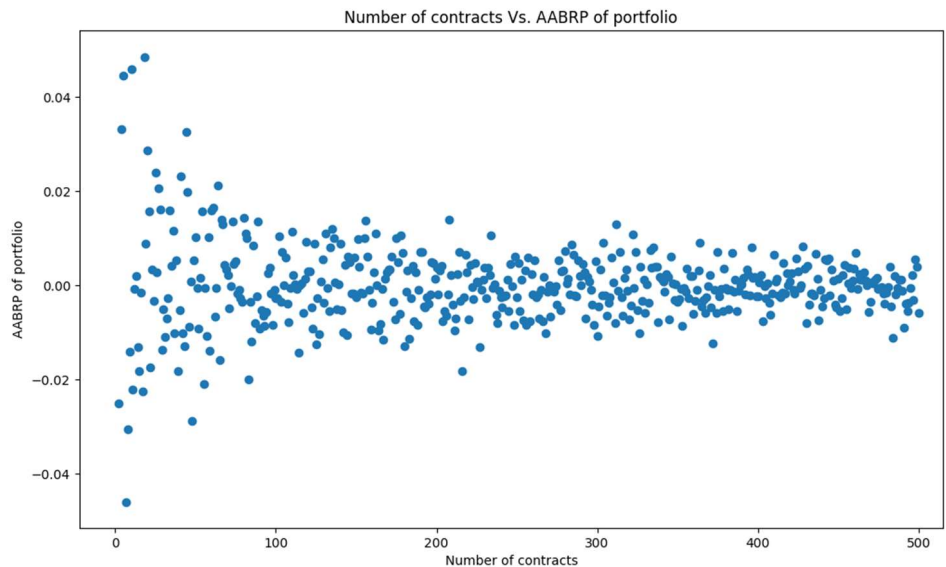

Figure 5(a) Relationship between the Number of Contracts in the Risk Portfolio and AABRP′

The convergence of AABRP′ with the increasing number of contracts is evident in two aspects: 1) As the number of contracts increases, the expected value of AABRP′ tends toward a constant value of 0; 2) As the number of contracts increases, the volatility of AABRP′ decreases. These conclusions are illustrated in Figure 5(b). Figure 5(b) shows the mean, standard deviation, maximum, and minimum of AABRP′ for every interval of 20 contracts, with the x-axis representing each incremental interval. As shown, the mean of AABRP′ consistently remains near zero, while the standard deviation of the portfolio basis risk decreases with the increase in the number of contracts.

To determine the correlation between the portfolio basis risk standard deviation and the number of contracts within the portfolio, different fitting tests were conducted. Figure 5(c) shows a nonlinear least squares fit testing the inverse proportional relationship between standard deviation and the number of contracts, resulting in parameters a = 0.482, b = 2.126, c = 0.003, with $R^2$ = 0.993.

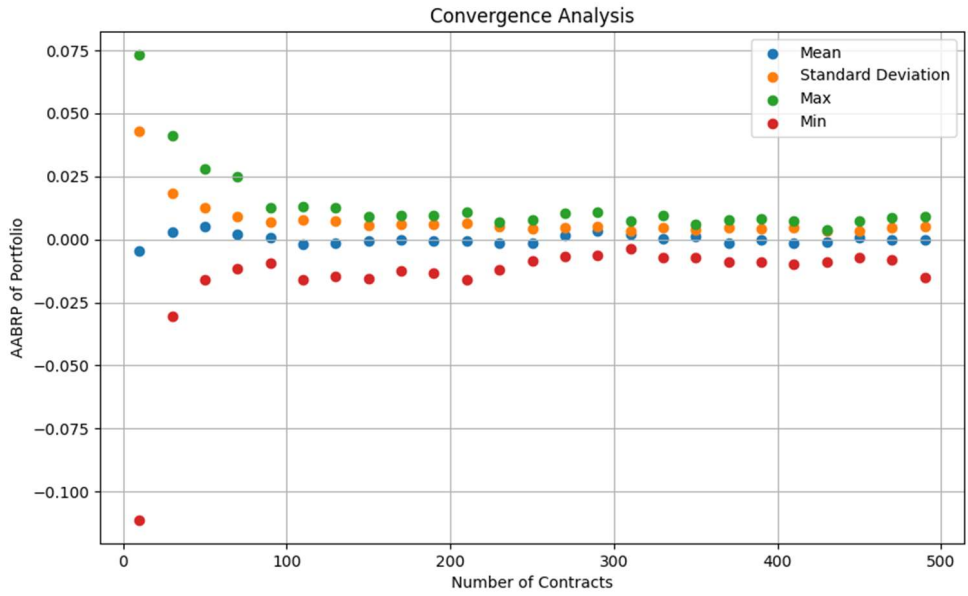

Figure 5(b) Variation of Mean, Extremes, and Standard Deviation of AABRP' in the Risk Portfolio with Changing Number of Contracts

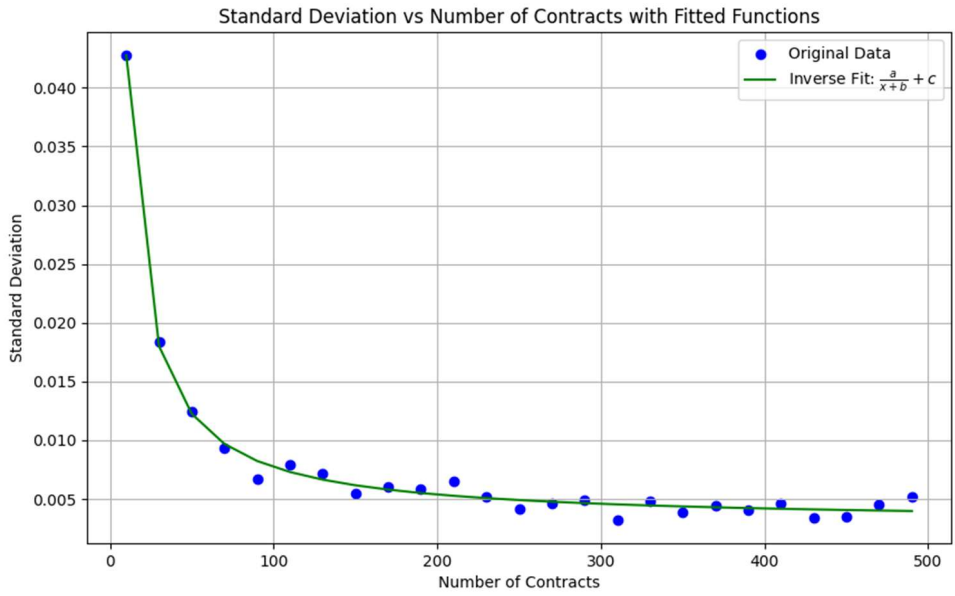

Figure 5(c) Fitted Curve of Standard Deviation of AABRP versus Number of Contracts

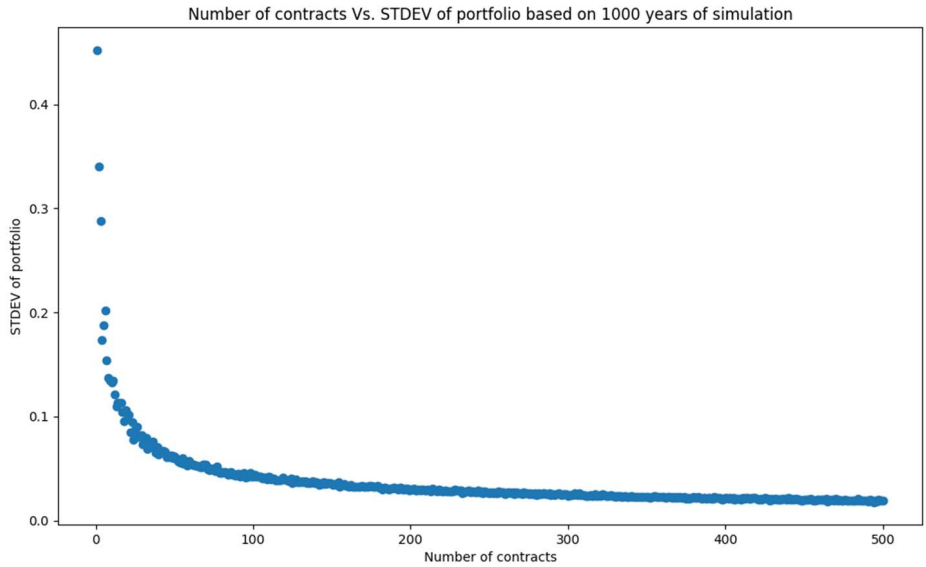

Figure 5(d) Relationship between the Number of Contracts in the Risk Portfolio and the Variance of Basis Risk

Figure 5(d) uses Pearson correlation, indicating a negative correlation of -0.611 and $R^2 = 0.373$ between the number of contracts and the volatility of the portfolio basis risk, statistically significant at the 99% confidence level. Figure 5(e) adopts an inverse proportional function model, with estimated parameters a = 1.967, b = 5.179, c = 0.019, with the model's $R^2 = 0.962$.

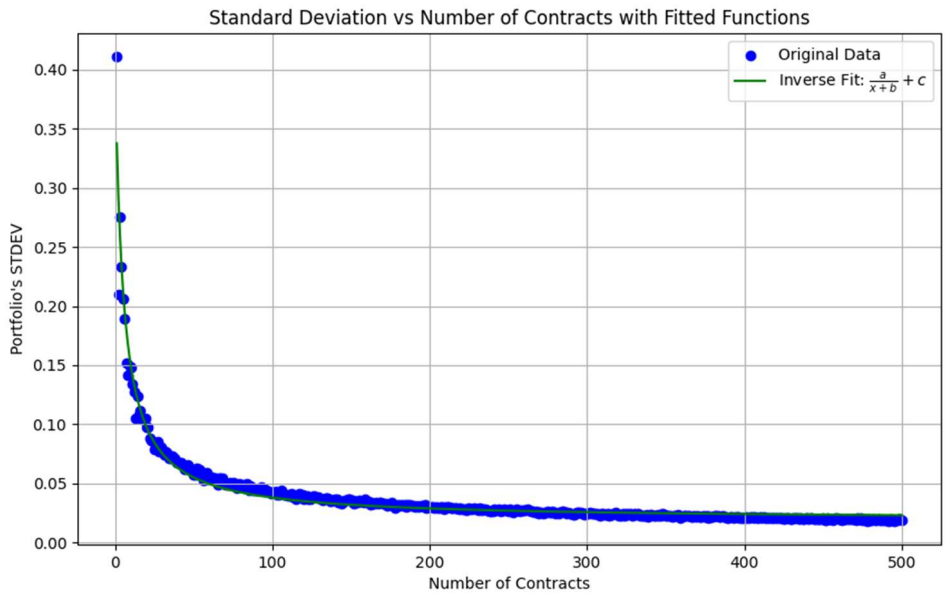

Figure 5(e) Fitted Curve of Basis Risk Standard Deviation versus Number of Contracts in the Risk Portfolio

## （2）The impact of spatial relationships among the insured exposure, reference weather station, and hazard event footprint radius on basis risk

For a single individual contract, we conducted 500 tests, each based on 1000 simulation years. For each of these 500 tests, the location of the insured exposure and reference weather station, as well as the hazard event footprint radius, remained unchanged throughout the 1000 simulation years. The location of the hazard event footprint and its severity changed randomly across the 1000 simulation years. Therefore, each test represents a unique spatial ratio which is defined as the distance from the exposure to the reference station and the event footprint radius.

Through 500 tests, 500 different AABRP values were generated for an analysis between AABRP and the spatial ratio mentioned above. As shown in Figure 6(a), the x-axis represents the spatial ratio between the distance from the insured object to the weather station relative to the event footprint radius. A larger ratio indicates that the distance between the insured exposure and the weather station is large, while the event footprint radius is small in comparison. A smaller ratio indicates that the insured exposure is relatively close to the weather station, and the event footprint radius is large in comparison. The y-axis represents AABRP. Figure 6(a) presents a spatial ratio range from 0 to over 140. To provide clearer insights, Figure 6(b) shows a zoomed-in section with a spatial ratio range on the x-axis from 0 to 20.

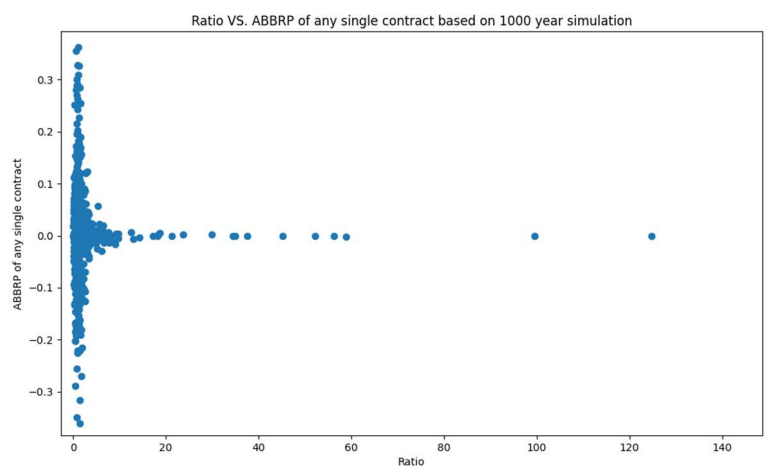

Figure 6(a) Relationship between Spatial Location and AABRP Based on 500 tests

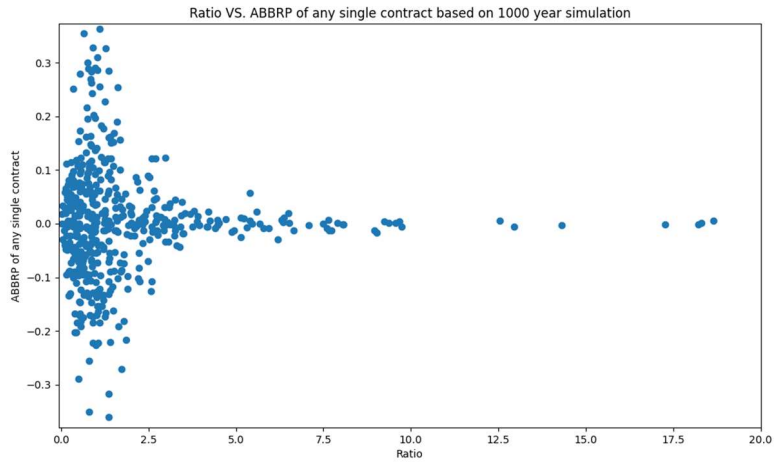

Figure 6(b) Relationship between Spatial Location and AABRP Based on 500 tests

(Zoom in section of spatial ratio from 0 to 20)

To examine the statistics of the results, the 500 tests are divided into spatial ratio incremental intervals of 0.05. The mean, standard deviation, maximum, and minimum AABRP values within each 0.05 spatial ratio interval are calculated, as shown in Figure 6(c). The x-axis represents the spatial ratio. Figure 6(d) is a zoomed-in section for spatial ratios between 0 and 10. As shown in the figures, mean values of AABRP appear to be symmetric around 0, increasing along the x-axis when the spatial ratio is below a certain threshold, and then decreasing along the x-axis and converging toward 0. Extrema of AABRP (including both maximum and minimum) exhibit similar characteristics: they are symmetric around 0, increase along the x-axis when the spatial ratio is below a certain threshold, and then start to converge to 0. Standard deviations shows similar pattern, initially increase, then decrease, and converge.

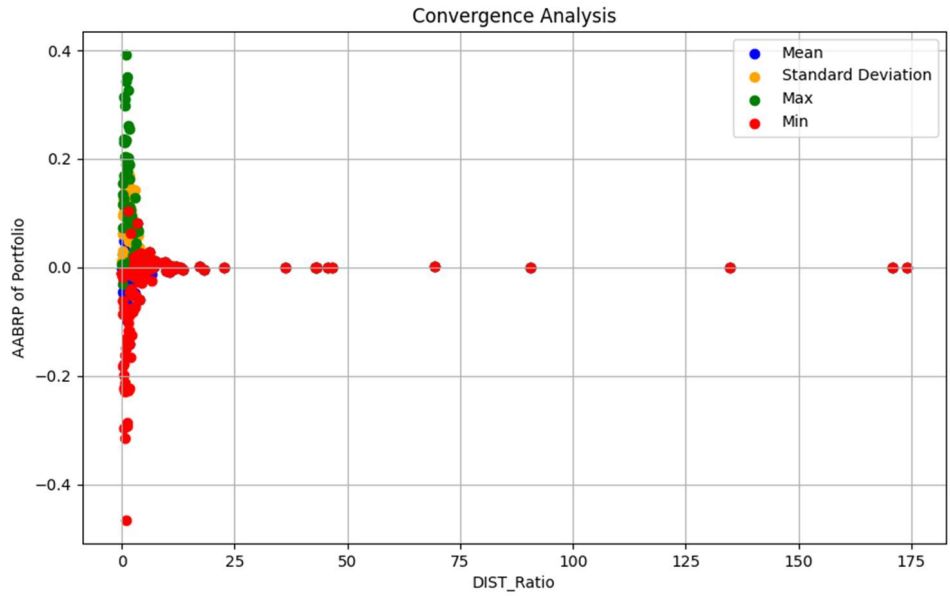

Figure 6(c) Variation of Mean, Extremes, and Standard Deviation of AABRP in the Risk Portfolio with Changes in Ratio

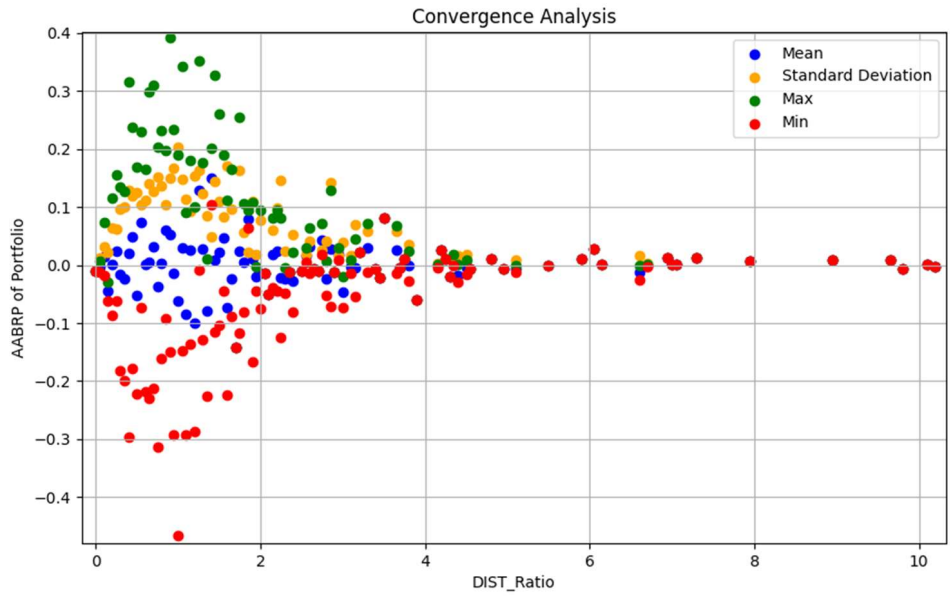

Figure 6(d) Variation of Mean, Extremes, and Standard Deviation of AABRP in the Risk Portfolio with Changes in Ratio (X-Axis Zoomed from 0 to 10)

To further examine the pattern, a "threshold regression" model is employed to estimate the threshold value where the transition from increasing to decreasing occurs (Hansen et al., 2000). In the first examination, threshold regression is applied to AABRP for individual tests among the 500 tests, yielding a threshold model regression result of 0.85, which is significant at the 99% confidence level. When the spatial ratio is lower than or equal to this threshold value, AABRP is positively correlated with the spatial ratio, with an R-square of 0.821, statistically significant at the 99% confidence level (p-value < 0.01). When the ratio is greater than the threshold, AABRP is negatively correlated with the spatial ratio, with an R-square of 0.388, statistically significant at the 99% confidence level (p-value < 0.01).

In the second examination, the threshold regression model is applied to the standard deviation of basis risk for any single test among the 500 tests. Figure 7 shows the relationship between the standard deviation of basis risk and the spatial ratio, ranging from 0 to 25 for a single individual parametric contract.

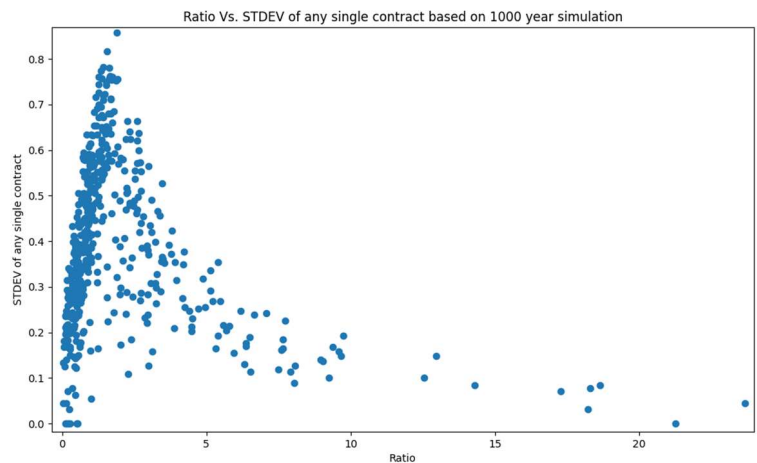

Figure 7(a) Relationship between Contract Basis Risk Variance and Spatial Ratio Based on 500 Simulations (X-Axis Zoomed from 0 to 25)

By applying the threshold regression model (Hansen et al., 2000), the threshold model regression result is 1.93, significant at the 99% confidence level. When the spatial ratio is lower than or equal to this threshold value, the standard deviation of basis risk is positively correlated with the spatial ratio, with an R-square of 0.597, statistically significant at the 99% confidence level (p-value < 0.01). When the spatial ratio is greater than this threshold, the standard deviation of basis risk is negatively correlated with the spatial ratio, with an R-square of 0.233, statistically significant at the 99% confidence level (p-value < 0.01).

To understand the transition of basis risks and basis risk volatilities around thresholds, we further analyzed the spatial relationship among insured exposure, reference weather station, and hazard event radius. The spatial ratio is defined as the distance between the insured exposure and the reference weather station relative to the hazard event radius. Therefore, in situations where the spatial ratio equals 0, the insured object and the weather station are approaching infinity relative to the hazard event footprint radius, or the event radius is infinitely large in comparison. In either case, this guarantees a 100% probability that both the exposure and station are within the hazard footprint, resulting in a basis risk and basis risk uncertainty of 0.

As the spatial ratio increases, the distance between the exposure and the station becomes larger, or the event footprint radius becomes smaller. In either scenario, the probability increases that the event footprint covers only the station or the exposure, but not both. Therefore, basis risk and basis risk volatility increase.

Beyond a certain threshold, when the distance between the station and the exposure is sufficiently large, or the event footprint radius is sufficiently small, the probability that the hazard event footprint covers neither the insured exposure nor the station increases, leading to a decrease in both basis risk and basis risk volatility.

In the scenario where the spatial ratio is extremely large, the distance between the station and the exposure is approaching infinitely large, or the event footprint radius is approaching to 0 in comparison. This ensures a 100% probability that the hazard event footprint covers neither the station nor the exposure, resulting in a basis risk and basis risk volatility of 0.

**(3) Correlation between hazard event severity and basis risk level**

Figure 9(a) shows the relationship between hazard event severity and AABRP for any single contract. When the hazard severity is below the contract parametric triggering threshold, basis risk is zero as no triggering occurs. When the severity exceeds the threshold, basis risk levels increase significantly, showing a generally symmetric random distribution around zero, with no significant statistical relationship with event severity. Figure 9(b) shows that the volatility of basis risk is zero when event severity is below the trigger threshold. When the severity exceeds the threshold, volatility increases and stabilizes, but without any significant statistical relationship with severity.

Therefore, the randomness of AABRP and standard deviation displayed in Figures 9(a) and 9(b) are primarily driven by spatial factors, which are decisive. On the other hand, there is no evidence showing that basis risk and its volatility increase along with an increase in severity.

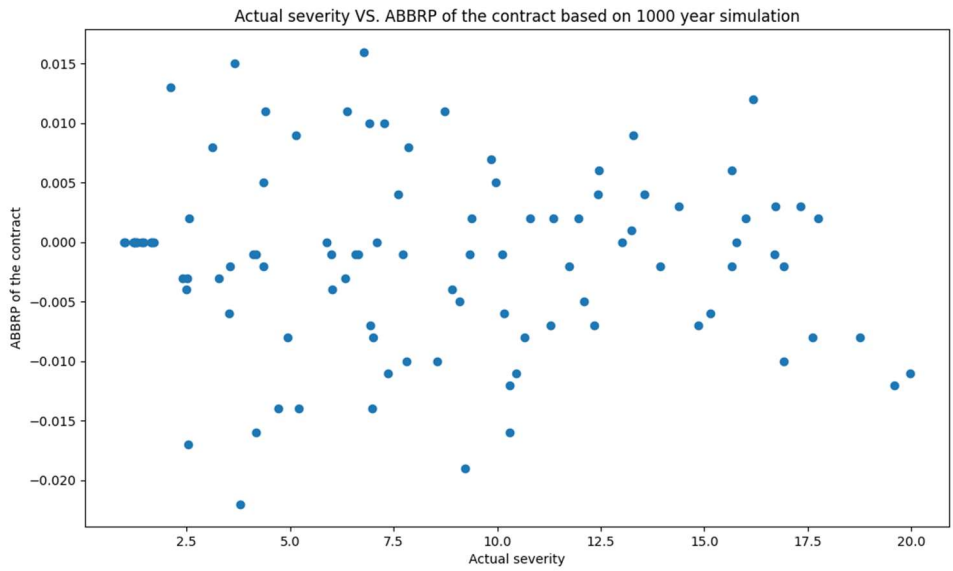

Figure 9(a) Relationship between Actual Severity and AABRP Based on 1000-Year Simulation

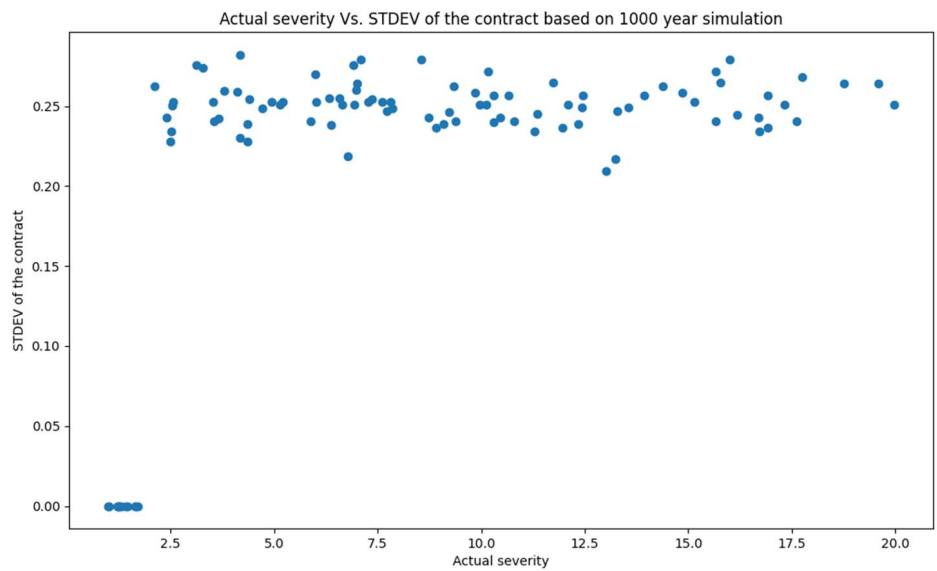

Figure 9(b) Relationship between Actual Severity and STDEV Based on 1000-Year Simulation

## Conclusion and Discussion

In this paper, we present an innovative study that employs Monte Carlo simulation methods to quantitatively analyze and evaluate basis risks and their statistical characteristics inherent in weather parametric insurance. To facilitate our study, we define the concept of the AABRP for the first time. The study systematically examines the effects of diversification on hedging and managing basis risk, as well as the correlations between basis risk and several key factors. The main findings are as follows:

(1) Diversification and Basis Risk: The overall basis risk level of a parametric insurance risk portfolio is strongly negatively correlated with its degree of diversification. As the number of independent weather parametric insurance contracts in the risk portfolio increases, the overall basis risk level decreases significantly. This is primarily due to the hedging effect of individual basis risks of independent insurance contracts. With increased diversification, this hedging effect becomes more pronounced, and the average basis risk approaches zero. Consequently, the uncertainty (standard deviation) of the overall basis risk also significantly decreases with increased diversification.

(2) Spatial Factors and Basis Risk: The spatial relationships among the insured exposure, the reference weather station, and the hazard event radius significantly impact the basis risk level of parametric insurance contracts. We define a spatial ratio as the ratio of the distance between the insured exposure and the weather station to the event radius. When the spatial ratio is lower than a certain threshold, basis risks and basis risk uncertainties positively correlated with the spatial ratio. Conversely, when the spatial ratio is greater than a certain threshold, the basis risk level and basis risk uncertainties negatively correlated with the spatial ratio.

(3) Impact of Event Severity: The impact of event severity on basis risk is relatively limited. When event severity is below the parametric trigger threshold, both basis risk and its uncertainty are zero. When severity exceeds the threshold, there is no significant correlation between event severity and the level or volatility of basis risk.

In the context of increasing climate change risks and the rapid development of weather parametric insurance, these findings provide valuable insights for insurance providers, policyholders, and financial institutions issuing or investing in parametric products, such including but not limited to parametric insurance-linked securities (ILS), options, derivatives, and other alternative products. (1) Basis risk is an inherent risk associated with parametric insurance. It represents the trade-off for a simplified claims settlement process and cannot be eliminated by technological solutions. (2) Hedging effectiveness may vary depending on hazard types and specific implementation conditions, but the inherent basis risk in weather parametric insurance can be effectively managed through portfolio diversification. Therefore, instead of limiting parametric insurance lines of business due to concerns about basis risks, bother insurer or insurerd,especially institutional policy holders, should consider expanding and forming highly diversified parametric insurance portfolios. This could be a crucial and effective solution for addressing basis risk. (3) The spatial relationship between the geographical location of the insured exposure and the reference weather station, as well as their ratio to the hazard event footprint size and scale, significantly affects both the level and volatility of parametric insurance basis risk. This should be given serious consideration in product design, such as the spatial distribution density of weather stations and the scale of hazard events (e.g., small footprint events like severe convective storms or large footprint events like tropical cyclones, or temperature-related events such as heat wave or cold wave which represents a large footprint with relatively uniform physical characteristics). (4) Hazard event severity appears to have no significant impact on basis risk. Therefore, parametric insurance can be effectively used to hedge high-severity catastrophe risks without concern that basis risk and its volatility will increase with hazard severity.

This study proposes a novel quantitative framework for analyzing the inherent basis risk in weather parametric insurance, with important empirical values. However, certain limitations exist, which, although do not affect the final conclusions, should be acknowledged: (1) The research is focused on weather parametric insurance. The scope could be expanded to include other types of perils. For instance, the same research methodology could be applied to geological risks with simple adjustments to the research assumptions. (2) In our study, insured exposure is represented by a point location. While this assumption is suitable for property exposure, such as a specific building, it may require refinement and calculation adjustments for agricultural insurance, which involves larger areas. (3) The spatial locations of the insured exposure, reference weather station, and event centroid are randomly defined within a [0,1] square space, and the disaster area is assumed to be circular. This shape factor might impact specific calculations. (4) For simplicity, hazard event severity is assumed to be uniform across the entire event footprint. These limitations will be addressed in future studies to enhance the robustness of the findings.